\documentclass[a4paper,11pt]{article}
\usepackage[english]{babel}
\usepackage[T1]{fontenc}
\usepackage{amsfonts,amsmath}
\usepackage{dsfont}
\usepackage{eepic}
\usepackage{graphicx}
\usepackage{url}

\newcommand{\harpoon}[1]{\overset{\rightharpoonup}{#1}}
\newcommand{\fatone}{\mathds{1}}
\newcommand{\ddt}{\frac{\operatorname{d}}{\operatorname{dt}}}

\renewcommand{\tilde}[1]{\widetilde{#1}}
\newcommand{\Rr}{\mathbb{R}}
\newcommand{\su}{\mathfrak{su}(2)}
\newcommand{\glnc}{\mathfrak{gl}(N,\mathbb{C})}
\newcommand{\sph}{\mathbb{S}^2}
\newcommand{\cote}{\operatorname{Cote}_x}
\newcommand{\grad}{\operatorname{grad}}
\newcommand{\cst}{\operatorname{const}}

\newcommand{\eg}{e.g. }
\newcommand{\ie}{i.e. }
\newcommand{\lhs}{l.h.s. }
\newcommand{\rhs}{r.h.s. }
\newcommand{\resp}{resp.}

\newcommand{\paraa}[1]{\big(#1\big)}
\newcommand{\parab}[1]{\Big(#1\Big)}
\newcommand{\parac}[1]{\bigg(#1\bigg)}

\renewcommand{\mid}{\mathds{1}}

\newcommand{\CCom}[3]{\Big[#1,\big[#2,#3\big]\Big]}
\newcommand{\Com}[2]{\big[#1,#2\big]}
\newcommand{\Acom}[2]{\big\{#1,#2\big\}}
\newcommand{\h}{\hbar}
\newcommand{\vphi}{\varphi}
\newcommand{\diag}{\operatorname{diag}}
\newcommand{\Wd}{W^\dagger}

\newcommand{\W}{W}
\newcommand{\Dt}{\tilde{D}}
\newcommand{\dt}{\tilde{d}}
\newcommand{\xv}{\vec{x}}

\newcommand{\xvn}{\vec{x}^{(n)}}

\newcommand{\cv}{\vec{c}}

\newcommand{\dn}{d^{(n)}}
\newcommand{\dtn}{\dt^{(n)}}

\numberwithin{equation}{section}

\title{\bf Fuzzy Riemann Surfaces} \author{Joakim Arnlind$^{\star}$,
Martin Bordemann$^{\sim}$, Laurent Hofer$^{\star\sim}$,\\Jens
Hoppe$^{\star}$ and Hidehiko Shimada$^{\dagger}$\\[1cm] {\small
\begin{minipage}{5cm} $\!\!^{\star}$Department of Mathematics\\ Royal
Institute of Technology\\ Lindstedtsv\"agen 25\\ S-10044 Stockholm.\\
\url{jarnlind@math.kth.se}\\ \url{hoppe@math.kth.se}
\end{minipage}
\begin{minipage}{5cm}
$\!\!\!^{\sim}$Laboratoire de MIA\\
4, rue des Fr\`eres Lumi\`ere\\
Universit\'e de Haute-Alsace\\
F-68093 Mulhouse.\\
\url{martin.bordemann@uha.fr}\\
\url{laurent.hofer@uha.fr}
\end{minipage}
\begin{minipage}{5cm}
$\!\!^{\dagger}$Max Planck Institute for\\
Gravitational Physics\\
(Albert Einstein Institute)\\
Am Mühlenberg 1\\
D-14476 Golm.\\
\url{hidehiko.shimada@aei.mpg.de}
\end{minipage}}\vspace{8mm}
}

\date{ }

\begin{document}

\maketitle

\begin{center}
  \begin{minipage}{13.5cm}
    \textsc{Abstract.} {\it We introduce C-Algebras (quantum analogues of
 compact Riemann surfaces), defined by polynomial relations in
 non-commutative variables and containing a real parameter that, when
 taken to zero, provides a classical non-linear, Poisson-bracket,
 obtainable from a single polynomial C(onstraint) function. For a
 continuous class of quartic constraints, we explicitly work out finite
 dimensional representations of the corresponding C-Algebras.}
  \end{minipage}
\end{center}

\section*{Introduction}

Harmonic homogenous polynomials in 3 commuting variables, upon
substitution of $N$-dimen\-sional representations of $\su$ for the
commuting variables, can be used to define a map from functions on
$\sph$ to $N\times N$ matrices, that sends Poisson brackets to matrix
commutators \cite{Hop82}. The result was dubbed ``Fuzzy Sphere'', in
\cite{Mad92}. In \cite{KL92} and \cite{BMS94} it was proven
(conjectured in \cite{BHSS91}) that the (complexified) Poisson algebra
of functions on \emph{any} Riemann surface arises as a
$N\rightarrow\infty$ limit of $\glnc$. Insight on how matrices can
encode topological information (certain sequences having been
indentifyable as converging to a particular function, but $\glnc$
lacking topological invariants) was gained in \cite{Shi04}.  That no
concrete anologue of the Fuzzy Sphere construction \cite{Hop82} for
higher genus compact surfaces could be found, and that the one found
for the torus \cite{FFZ89,Hop88} was of a very different nature,
remained an unsolved puzzle, just as the rigidity of the 2
constructions.  We are happy to announce a resolution by presenting a
unified (\emph{and} concrete, as well as non-rigid, and intuitively
simple) treatment for general compact Riemann surfaces.  In section
\ref{sec:genus_g_desc} we describe Riemann surfaces of genus $g$
embedded in $\Rr^3$ as inverse images of polynomial
contraint-functions, $C(\vec{x})$. In section \ref{sec:pois_brack_r3}
we define a Poisson bracket on $\Rr^3$, to be restricted to the
embedded Riemann surface. In section \ref{sec:genus_0_quant} we review
the quantization for the round 2-sphere. In section
\ref{sec:genus_g_const} we outline our construction for general
genus. In section \ref{sec:torus_const} we work out the conjectured
construction for a continuous class of tori and deformed spheres. In
section \ref{sec:sing} we discuss how the classical singularity at
$\mu=1$ is reflected in the quantum world.

\section{Genus $g$ Riemann surfaces}\label{sec:genus_g_desc}

\noindent The aim of this section is to present compact connected
Riemann surfaces of any genus embedded in $\Rr^3$ by inverse images of
polynomials. For this purpose we use the regular value theorem and
Morse theory. Let $C$ be a polynomial in 3 variables and define
$\Sigma=C^{-1}(\{0\})$. What are the conditions on $C$, for $\Sigma$
to be a genus $g$ Riemann surface? If $C$ is a submersion on $\Sigma$,
then $\Sigma$ is an orientable submanifold of $\Rr^3$. $\Sigma$ has to
be compact and of the desired genus. For further details see
\cite{Hir76,Hof02}.

\paragraph{}
The classification of 2 dimensional compact (connected) manifolds is
well known. In this case, there is a one to one correspondance between
topological and diffeomorphism classes. The result is that any compact
orientable surfaces is homeomorphic (hence diffeomorphic) to a sphere
or to a surface obtained by glueing tori together (connected sum). The
number $g$ of tori is called the genus and is related to the
Euler-Poincar\'{e} characteristic by the formula $\chi=2-2g$.

\paragraph{Morse theory}
To compute $\chi(\Sigma)$ we apply Morse theory to a specific
function. A point $p$ of a (smooth) function $f$ on $\Sigma$ is a
singular point if $Df_p=0$, in which case $f(p)$ is a singular
value. At any singular point $p$ one can consider the second
derivative $D^2 f_p$ of $f$ and $p$ is said to be non-degenerate if
$\det(D^2 f_p)\neq 0$. Moreover one can attach an index to each such
point depending on the signature of $D^2 f$: 0 if positive, 1 if
hyperbolic and 2 if negative. A Morse function is a function such that
every singular point is non-degenerated and singular values all
distinct. Then $\chi(\Sigma)$ is given by the formula:
$$ \chi(\Sigma)=n(0)-n(1)+n(2), $$ where $n(i)$ is the number of
singular points which have an index $i$.

\paragraph{}
The $\cote$ function is defined as the restriction of the first
projection on the surface. It's not necessarily a Morse function (one
has to choose a ``good'' embedding for that), but the singular points
are those for which the gradient $\grad C$ is parallel to the $Ox$
axis. Moreover the Hessian matrix of $\cote$ at such a point $p$ is:
$$ -\frac{1}{\frac{\partial C}{\partial x}(p)}\left(
   \begin{matrix}
      \frac{\partial^2 C}{\partial y^2}(p) & \frac{\partial^2 C}{\partial y\partial z}(p)\\
      \frac{\partial^2 C}{\partial y\partial z}(p) & \frac{\partial^2 C}{\partial z^2}(p)
   \end{matrix}\right). $$

\paragraph{Polynomial model}

Take $$ C(\vec{x})=(P(x)+y^2)^2+z^2-\mu^2, $$ where $\mu>0$,
$P(x)=a_{2k} x^{2k}+a_{2k-1} x^{2k-1}+\cdots+a_1 x+a_0$ with
$a_{2k}>0$ and $k>0$. Obviously $\Sigma$ is closed and bounded (even
degree of $P$) hence compact. $\Sigma$ is a submanifold of $\Rr^3$ if,
and only if for each $p\in \Sigma$, $DC_p\neq 0$ which is equivalent
to requiring that the polynomials $P-\mu$ and $P+\mu$ have only simple
roots. The singular points of the $\cote$ function on $\Sigma$ are the
points $(x,0,0)$ such that $P(x)^2=\mu^2$ and the Hessian matrix is:
$$ -\frac{1}{\frac{\partial C}{\partial
x}(x,0,0)}\left(\begin{matrix}4P(x) & 0 \\ 0 &
2\end{matrix}\right). $$ Hence it is positive or negative if, and only
if $P(x)=\mu$ and hyperbolic if, and only if $P(x)=-\mu$. With the
fact that $P(x)$ can't be zero at a singular point, it also proofs
that $\cote$ is a genuine Morse function. Finally,
$$ n(0)+n(2)=\#\{ P = \mu \}\qquad\text{and}\qquad n(1)=\#\{ P = -\mu
\}. $$ If the polynomial $P-\mu$ has exactly 2 simple roots and the
polynomial $P+\mu$ has exactly $2g$ simple roots, then
$\chi(\Sigma)=2-2g$ and $\Sigma$ is a surface of genus $g$.

\paragraph{Explicit construction of $P$}

Let $g>0$. Set:
\begin{tabbing}
   \hspace{5mm}\=(i)\hspace{8mm}\=$G(t)=(t-1)(t-2^2)\ldots(t-g^2)$\hspace{1cm}\=and\hspace{1cm}\=$M=\underset{0\leq
   t\leq g^2+1}{\max} G(t),\quad \alpha\in\ \left(0,\frac{2\mu}{M}\right)$\\[2mm]
 \>(ii)\>$Q(x)=\alpha
   G(x)-\mu$\>and\>$P(x)=Q(x^2)$
\end{tabbing}
One can directly see that $Q+\mu$ has exactly $g$ simple roots, hence
$P+\mu$ has exactly $2g$ simple roots. For $t\in [0;g^2+1]$, the
function $Q(t)-\mu$ has no zero. On the other hand, for $t\geq g^2+1$,
$Q(t)-\mu$ is strictly growing and has exacly one zero. Consequently
the polynomial $P-\mu$ has exactly 2 simple roots and the surface
$\Sigma$ defined above is a genus $g$ compact Riemann surface.

\section{Poisson brackets in $\Rr^3$}\label{sec:pois_brack_r3}

For arbitrary $C:\Rr^3\longrightarrow\Rr$ (twice continuously differentiable)
\begin{equation}
  \left\lbrace f,g \right\rbrace_{\Rr^3} := \vec\nabla C\cdot(\vec\nabla f\times \vec\nabla g)\label{eq:pois_1}
\end{equation}
defines a Poisson bracket for functions on $\Rr^3$ (see \eg Nowak
\cite{Now97} who studied the formal deformability of
(\ref{eq:pois_1}))\footnote{While we did not (yet find a way to) use his results,
we are very grateful for his ``New Year's Eve'' explanations, as well as
providing us with his Ph.D. Thesis.}.
Let $\Sigma_g\subset\Rr^3$ be described, as in section 1,
by a ``constraint'':
\begin{equation}
  \frac{1}{2}C(\vec x):=\psi(x,y)+\frac{z^2-1}{2}\overset{!}{=}0.\label{eq:pois_2}
\end{equation}
The Poisson brackets between $x$,$y$ and $z$ then read:
\begin{eqnarray}
\{ x,y \}_{\Rr^3} & = & z\nonumber\\
\{ y,z \}_{\Rr^3} & = &\psi_x\nonumber\\
\{ z,x \}_{\Rr^3} & = & \psi_y\label{eq:pois_3}.
\end{eqnarray}
Explicitely, substituting $\{x,y\}$ for $z$, one obtains
\begin{equation}
  \psi(x,y)+\frac{1}{2}\{x,y\}_{\Rr^3}^2 = \cst\left(=\frac{1}{2}\right),\label{eq:pois_4}
\end{equation}
\resp
\begin{eqnarray}
  \psi_x & = & \big\{y,\{x,y\}_{\Rr^3}\big\}_{\Rr^3}\nonumber\\
  \psi_y & = & \big\{\{x,y\}_{\Rr^3},x\big\}_{\Rr^3}.\label{eq:pois_5}
\end{eqnarray}
Let $x(\sigma_1,\sigma_2)$, $y(\sigma_1,\sigma_2)$,
$z(\sigma_1,\sigma_2)$ be a local parametrisation of
$\Sigma_g$. Restricting $\left\lbrace f,g \right\rbrace_{\Rr^3}$ to a
Poisson bracket, $\left\lbrace f,g \right\rbrace_C$, on the surface
$C(\harpoon x)=0$, and realizing $\{f,g\}_C$ on $\Sigma_g$ locally as
$$\frac{1}{\rho(\sigma^1,\sigma^2)}\left(\frac{\partial
f}{\partial\sigma^1}\frac{\partial g}{\partial\sigma^2}-\frac{\partial
g}{\partial\sigma^1}\frac{\partial f}{\partial\sigma^2}\right),$$ the
relation (equivalence!, up to different constant values on the \rhs of
(\ref{eq:pois_4})) between (\ref{eq:pois_4}) and (\ref{eq:pois_5}) is
seen as follows: differentiating (\ref{eq:pois_4}) with respect to the
local parameters, $\varphi^1$ and $\varphi^2$, one obtains a linear
system of equations for $\psi_x$ and $\psi_y$, whose algebraic
solution (via Cramers rule, \eg) gives (\ref{eq:pois_5}).  To go from
(\ref{eq:pois_5}) to (\ref{eq:pois_4}) (with the constant unspecified,
of course) one either notes simply that the \lhs of (\ref{eq:pois_4})
commutes with both $x$ and $y$ (according to (\ref{eq:pois_5})), or
one directly solves (\ref{eq:pois_5}) via a hodograph-transformation
(cp. \cite{BKL05}, in which Poisson bracket equations are considered,
whose solutions also contain surfaces of general type\footnote{Apart from
spheres, however, these surfaces are either non-polynomial, or
non-compact -- or both -- causing the corresponding quantum-algebras to
be necessarily \emph{different} from ours.});
changing independent variables from $\sigma^1$, $\sigma^2$ to
$x_1=x(\sigma^1,\sigma^2)$ and $x_2=y(\sigma^1,\sigma^2)$; using
(deriving) (as e.g. in \cite{BH93})
\begin{eqnarray*}
  \{x,y\}_C & =: & J(x_1,x_2)\\
  \{f,x\}_C & = & -Jf_y\\
  \{f,y\}_C & = & Jf_x,
\end{eqnarray*}
(\ref{eq:pois_5}) then becomes
\begin{eqnarray*}
  -JJ_x & = & \psi_x\\
  -JJ_y & = & \psi_y,
\end{eqnarray*}
\ie $\frac{1}{2} J^2+\psi=\cst$.

\section{``The fuzzy sphere'' \cite{Hop82}}\label{sec:genus_0_quant}

Consider the usual spherical harmonics,
$$ \big\{
Y_{lm}(\theta,\varphi)\big\}_{\overset{l=1,\ldots,\infty}{\underset{m=-l,\ldots,+l}{}}},
$$ eigenfunctions of the Laplace operator on $\sph$ ($\triangle_{\sph}
Y_{lm} = -l(l+1) Y_{lm}$). Write them as harmonic homogenous
polynomials in $x_1=r\sin\theta\cos\varphi$,
$x_2=r\sin\theta\sin\varphi$ and $x_3=r\cos\theta$ (restricted to
$r^2=\vec{x}^2=1$):
\begin{equation}\label{eq:fuzz_1}
Y_{lm}(\theta,\varphi) = \sum c_{a_1 a_2 \cdots a_l}^{(m)}\ x_{a_1} x_{a_2} \cdots x_{a_l}
\end{equation}
(where the tensor $c_{\ldots}$ is by definition traceless and totally
symmetric), and then replace the commuting variables $x_a$ by
generators $X_a$ of the $N$-dimensional irreducible (spin
$s=\frac{N-1}{2}$) represensation of $\su$, to obtain $N^2-1$ $N\times
N$-matrices:
\begin{equation}
T_{lm}^{(N)} :=\gamma_{Nl} \sum c_{a_1 a_2 \cdots a_l}^{(m)}\ X_{a_1}
X_{a_2} \cdots X_{a_l}\qquad\text{for}\quad l=1,\ldots, N-1 \quad
m=-l,\ldots,+l\,;\label{eq:fuzz_2}
\end{equation}
automatically, $T_{lm}^{(N)} \equiv 0$ for $l\geq N$.  Instead of
having $\vec{X\,}^2:=X_1^2+X_2^2+X_3^2$ equal to
$\frac{N^2-1}{4}\fatone$ (the usual normalisation), it is advantageous
to choose the normalisation $\vec{X}^2=\fatone$,
\begin{equation}\label{eq:fuzz_3}
[X_a,X_b]=\frac{2 i}{\sqrt{N^2-1}} \epsilon_{abc} X_c,
\end{equation}
and then $\gamma_{Nl} = -i\sqrt{\frac{N^2-1}{4}}$.
As the Poisson bracket on $\sph$,
\begin{equation}\label{eq:fuzz_5}
\{f,g\}_{\sph}(\theta,\varphi) :=
\frac{1}{\sin\theta}\left(\frac{\partial
f}{\partial\theta}\frac{\partial g}{\partial\varphi}-\frac{\partial
g}{\partial\theta}\frac{\partial f}{\partial\varphi}\right)
\end{equation}
can be obtained by restricting the Poisson bracket
\begin{equation}
\{f,g\}_{\Rr^3}(\vec{x}) := \vec{x}\cdot\left(\vec\nabla f(\vec{x})\times\vec\nabla g(\vec{x})\right)\label{eq:fuzz_6}
\end{equation}
to $\sph$ (via $\vec{x\,}^2=1$), $\{Y_{lm},Y_{l'm'}\}_{\sph}$ can be
computed from
\begin{equation}\label{eq:fuzz_7}
\left\{r^l Y_{lm},r^{l'} Y_{l'm'}\right\}_{\Rr^3} = \sum c_{a_1\ldots
a_l}^{(m)}\ c_{b_1\ldots b_{l'}}^{(m')}\ \left\{x_{a_1} x_{a_2} \cdots
x_{a_l}, x_{b_1} x_{b_2} \cdots x_{b_{l'}}\right\}
\end{equation}
by using the derivation property, and
\begin{equation}\label{eq:fuzz_8}
\{x_a,x_b\} = \epsilon_{abc} x_c
\end{equation}
(following from (\ref{eq:fuzz_6})), as well as then decomposing the
resulting polynomial of degree $l+l'-1$ into harmonic homogenous ones,
to obtain the structure constants of the Lie-Poisson algebra of
functions on the $2$-sphere (in the basis of the spherical
harmonics).\\ Calculating
\begin{equation}\label{eq:fuzz_9}
\left[T_{lm}^{(N)},T_{l'm'}^{(N)}\right] = -\frac{N^2-1}{4} \sum
c_{a_1\ldots a_l}^{(m)}\ c_{b_1\ldots b_{l'}}^{(m')}\ \left[X_{a_1}
X_{a_2} \cdots X_{a_l}, X_{b_1} X_{b_2} \cdots X_{b_{l'}}\right]
\end{equation}
the first step is identical to the one after (\ref{eq:fuzz_7}), while
any further use of the commutation relations (\ref{eq:fuzz_3}) --
necessary to obtain the desired traceless totally symmetric tensors --
induces factors of $1/\sqrt{N^2-1}$; hence one finds agreement to
leading order of $N$ of the structure constants of $\glnc$, in the
basis $\left\{T_{lm}^{(N)}\right\}_{l=1,\ldots,N-1}$, satisfying 
$$ \operatorname{Tr}\left(T^{(N)\dagger}_{lm} T^{(N)}_{l'm'}\right) =
\delta_{ll'}\delta_{mm'}\frac{(N+l)!}{16\pi(N-1-l)!(N^2-1)^{l-1}},$$
with those of the Poisson algebra.

\section{The construction for general Riemann surfaces}\label{sec:genus_g_const}

Let us consider compact Riemann surfaces $\Sigma_g\in\Rr^3$ described by
\begin{equation}\label{eq:genus_g_1}
\left(P(x)+y^2\right)^2+z^2=\cst\left(=1\right)
\end{equation}
(with P, as in section (\ref{sec:genus_g_desc}), an even polynomial of degree $2g$), \resp
\begin{eqnarray}
\big\{y,\{x,y\}\big\} & = & P'(x)(P(x)+y^2)\nonumber\\
\big\{\{x,y\},x\big\} & = & 2y(P(x)+y^2)\label{eq:genus_g_2}
\end{eqnarray}
(for (\ref{eq:pois_3})). We claim that fuzzy analogues of $\Sigma_g$
can be obtained via matrix analogues of (\ref{eq:genus_g_1}) and
(\ref{eq:genus_g_2}). Apart from possible ``explicit $1/N$
corrections'', direct ordering questions arise both on the \rhs of
(\ref{eq:genus_g_2}), and in (\ref{eq:genus_g_1}), while on the \lhs
of (\ref{eq:genus_g_2}) one replaces Poisson brackets by
$\frac{1}{i\hbar}$(commutator(s)).\\ Consider therefore the problem of
looking for matrices $X,Y$ satisfying
\begin{equation}
(P(X)+Y^2)^2-\frac{1}{\hbar^2}[X,Y]^2=\fatone,\label{eq:genus_g_3}
\end{equation}
\resp
\begin{eqnarray}
\frac{1}{\hbar^2}\big[X,[X,Y]\big] & = & 2Y^3+YP(X)+P(X)Y\label{eq:genus_g_4}\\
\frac{1}{\hbar^2}\big[Y,[Y,X]\big] & = & \sum_{r=1}^{2g-1} a_r \sum_{i=0}^{r-1} X^i\left(P(X)+Y^2\right)X^{r-1-i}\label{eq:genus_g_5}
\end{eqnarray}
if $P(X)=\sum_{r=0}^{2g} a_r X^r$; the \rhs of (\ref{eq:genus_g_5})
will also be denoted by $P(X)'|_{\varphi=P(X)+Y^2}$ (for a term $X^4$ in $P(X)$, \eg, $P'(X)|_{\varphi}$
would correspondingly include $X^3\varphi+X^2\varphi X+X\varphi
X^2+\varphi X^3$). This ordering in (\ref{eq:genus_g_4}) and
(\ref{eq:genus_g_5}) is consistent, as
\begin{eqnarray}
\big([X,Y]Y\big)X-[X,Y](YX) & = & \Big(Y[X,Y]+\big[[X,Y],Y\big]\Big)X-[X,Y]\big(XY+[Y,X]\big)\label{eq:genus_g_6}\\
 & = & \cdots \nonumber\\
 & = & \Big[Y,\big[[X,Y],X\big]\Big]+\Big[\big[[X,Y],Y\big],X\Big]
\end{eqnarray}
indeed equals to zero (insert (\ref{eq:genus_g_4}) and
(\ref{eq:genus_g_5}) for the 2 double commutators to get $P(X)Y^2-Y^2
P(X)+[P'(X)|_{Y^2},X]=\cdots=0$), which is has to, due to
associativity of matrix multiplication (and resulting Jacobi
identity). 

Finding (for specific values of $\hbar^2$) concrete representations of
(\ref{eq:genus_g_4}) and (\ref{eq:genus_g_5}), \resp
(\ref{eq:genus_g_3}), let alone classifying them, is of course a very
complicated task. We succeeded in doing so for $P(x)=x^2-\mu$
(corresponding to a torus when $\mu>1$, and deformed spheres, when
$-1<\mu<1$, see section (\ref{sec:torus_const}), but first we would
like to outline some qualitative features involved for the general
case. As mentioned in the introduction, one of the
puzzles was the rigidity of the construction for the round 2-sphere
\cite{Hop82} and the toral rational rotation algebra \cite{FFZ89} (see
also \cite{Hop89}). In the present construction we now have a
(generally, \ie apart from certain critical values -- signaling
topology change) continuous dependence on the data ($P$) which
describe the Riemann surface. For given $P$, and $\hbar$, we define
the corresponding Fuzzy Riemann Surface $\Sigma_\hbar (P)$ (for fixed
$P$, expected to exist for infinitely many discrete values of $\hbar$,
coming arbitrarily close to zero) as the algebra generated by the
corresponding finite ($N$-)dimensional solutions $(X,Y)$ of
(\ref{eq:genus_g_4}),(\ref{eq:genus_g_5}), \resp
(\ref{eq:genus_g_3}). According to the semiclassical philosophy
explained below (cp. \cite{Shi04}) $X$ and $Y$ will exhibit eigenvalue
sequencies (generically smoothly depending on $P$) characteristic of
the topological type, reflecting the behaviour of the corresponding
classical embedding functions $x$ and $y$.

\paragraph{}

Let us observe that we can read off information of topology from a
generic single function $f$ by using Morse theory. This Morse
theoretic information of topology manifests itself in the eigenvalue
distribution of the matrix $\hat{f}$ corresponding to the function
$f$:\\ the key idea is to introduce an auxilary Hamiltonian dynamical
system, whose phase space is the surface and whose Hamiltonian is
given by $f$. Thus we consider ordinary differential equations
\begin{eqnarray}
\ddt \sigma^1 =\{\sigma^1,f \},\nonumber\\
\ddt \sigma^2 =\{\sigma^2,f \},\label{eq:genus_g_7}
\end{eqnarray}
where $\sigma^1, \sigma^2$ parametrise the embedded surface. Since
classical orbits of (\ref{eq:genus_g_7}) are equal-$f$ lines on the
surface, the family of the classical orbits exhibits branching
processes which exactly reflect Morse theoretic information. The
eigenvalue distribution of $\hat{f}$ is determined (to leading order
in $\frac{1}{N}$) by the Bohr-Sommerfeld rule. That is, the
eigenvalues are values of $f$ on classical orbits which are such that
the area between adjacent orbits is equal to the total area of the
surface divided by $N$.  It follows that the eigenvalues are grouped
into subsets each corresponding to the branches on the surface. These
subsets are called in \cite{Shi04} as eigenvalue sequences, and have
the property that (for sufficiently large $N$) the eigenvalues
belonging to a sequence rise in a smooth way. The branching processes
of the sequences are the same as those of the classical
orbits. Furthermore, using the eigenvalue sequences and
(\ref{eq:genus_g_7}), a rule to calculate general (off-diagonal)
matrix elements of a matrix $\hat{g}$ corresponding to a general
function $g$ was given in \cite{Shi04}.  Essential properties of the
matrix regularisation, such as the correspondence of the matrix
commutator and the Poisson bracket, can be derived from those rules.

\section{Explicit solutions for tori and deformed spheres:\\
Representations of the simplest non-linear $C$--Algebras}\label{sec:torus_const}

Let now $P(x)=x^2-\mu$ which, for $\mu>1$ describes a torus, and for
$-1<\mu<1$ a (deformed) sphere. We will construct solutions for
the corresponding matrix equations, which we take as
\begin{align}
  &[X,Y] = i\h Z\label{XY}\\
  &[Y,Z] = i\h\Acom{X}{X^2+Y^2-\mu}\label{YZ}\\
  &[Z,X] = i\h\Acom{Y}{X^2+Y^2-\mu}\label{ZX}\\
  &\parab{X^2+Y^2-\mu}^2+Z^2=\mid\label{C}
\end{align}
(in this section, $\Acom{A}{B}$ denotes the anti-commutator $AB+BA$,
and \emph{not} a Poisson-bracket).
Denoting $X^2+Y^2-\mu$ by $\varphi$, one finds (using the Jacobi-identity,
the derivation property, and equations \eqref{YZ} + \eqref{ZX}) that 
\begin{equation}
  \begin{split}
    \Com{\vphi}{Z} &= \Acom{X}{\Com{X}{Z}}+\Acom{Y}{\Com{Y}{Z}}\\
    &=-i\h \Acom{X}{\Acom{Y}{\vphi}}+i\h \Acom{Y}{\Acom{X}{\vphi}}\\
    &=i\h\CCom{\vphi}{X}{Y} = -\h^2\Com{\vphi}{Z},
  \end{split}
\end{equation}
hence $\vphi(X,Y)$ and $Z$ commute, and can be diagonolized
simultaneously; it then also follows that $\vphi^2+Z^2$ is central,
i.e. commutes with $X,Y$ and $Z$.

In complex notation, $\W:=X+iY$, \eqref{YZ} and \eqref{ZX} can be
written together as
\begin{align}
  \parab{\W^2\Wd+\Wd\W^2}\paraa{\h^2+1}=
  4\mu\h^2\W + 2(1-\h^2)\W\Wd\W,\label{W}
\end{align}
from which the crucial commutativity of $D:=\W\Wd$ and $\Dt:=\Wd\W$,
\begin{equation}
  \begin{split}
    \Com{\W\Wd}{\Wd\W}
    &=\Com{X^2+Y^2-i[X,Y]}{X^2+Y^2+i[X,Y]}\\
    &=2i\CCom{X^2+Y^2}{X}{Y}=0,
  \end{split}
\end{equation}
also follows directly (by using \eqref{W}). In the basis where both
$D$ and $\Dt$ are diagonal
\begin{align*}
  D &= \diag(d_1,\ldots,d_N)\\
  \Dt &= \diag(\dt_1,\ldots,\dt_N),
\end{align*}
\eqref{W} becomes 
\begin{align}
  \W_{ij}\parac{\paraa{\h^2+1}(\dt_i+d_j)+\paraa{\h^2-1}(d_i+\dt_j)-4\mu\h^2}=0\label{WD},
\end{align}
and from $(\W\Wd)\W=\W(\Wd\W)$ we obtain
\begin{align}
  \W_{ij}\paraa{d_i-\dt_j}=0\label{WDtwo}
\end{align}
(which in fact was already used when writing \eqref{WD}
``symmetrically''). Thus, for $\W_{ij}\neq 0$, \eqref{WD} and
\eqref{WDtwo} are equivalent to
\begin{equation}
  \begin{split}
  &\begin{pmatrix}
    d_j \\ \dt_j
  \end{pmatrix}
  =
  \begin{pmatrix}
    \alpha & -1 \\ 1 & 0
  \end{pmatrix}
  \begin{pmatrix}
    d_i \\ \dt_i
  \end{pmatrix}
  +
  \begin{pmatrix}
    \delta \\ 0
  \end{pmatrix}\\
  &\alpha=2\frac{1-\h^2}{1+\h^2}=2\cos 2\theta\quad,\quad
  \delta = 4\mu\frac{\h^2}{1+\h^2}=4\mu\sin^2\theta,
  \end{split}\label{Matrixiteration}
\end{equation}
where we have put $\h=\tan\theta$. \eqref{Matrixiteration} is of the form
\begin{align}
  \xv_j = A\xv_i + \cv\quad;\label{Ax}
\end{align}
so, for
\begin{align}
  W &= 
  \begin{pmatrix}
    0 & w_1 & 0   & \cdots & 0 \\
    0 & 0   & w_2 & \cdots & 0 \\
    \vdots & \vdots & \ddots & \ddots & \vdots \\
    0 & 0 & \cdots & 0 & w_{N-1} \\
    w_N & 0 & \cdots & 0 & 0
  \end{pmatrix},\quad w_k\neq 0,\\
  \xv^{(n+1)} &= A\xvn+\vec{c},\quad n=1,2,\ldots,N\label{xvn}
\end{align}
are the only equations relating the $N$ eigenvalue pairs.
What can we say about the possibility of coming back to
the same point $\xv$ after $N$ iterations, i.e 
\begin{align}
  \xv^{(N+1)}-\xv=A^N\xv+ \parab{\mid+A+A^2+\cdots + A^{N-1}}\cv-\xv=0.\label{ANit}
\end{align}
Multiplying by $(\mid-A)$ gives
\begin{align}
  \paraa{\mid-A^N}\parab{(\mid-A)\xv+\cv}=0,
\end{align}
and one deduces that either $\det\paraa{\mid-A^N}\neq 0$, in which case 
\begin{align}
  \xv=(\mid-A)^{-1}\cv=\frac{1}{2-\alpha}
  \begin{pmatrix}
    \delta \\ \delta
  \end{pmatrix}
  =
  \begin{pmatrix}
    \mu \\ \mu
  \end{pmatrix},
\end{align}
-- a fix point of the transformation (i.e.
$A\begin{pmatrix}\mu\\ \mu\end{pmatrix}+\cv=\begin{pmatrix}\mu\\
\mu\end{pmatrix}$, making all $d$'s and $\dt$'s equal to one another,
i.e. $Z\equiv 0$, which we don't want) -- or $2N\theta$ is a multiple
of $2\pi$, i.e. 
\begin{align}
  \h = \tan\parac{\frac{\pi}{N}k},
\end{align}
in which case \eqref{ANit} holds for \emph{every} $\xv$, as then
$\mid+A+\cdots+A^{N-1}=0$ and $A^N=\mid$ (the eigenvalues of $A$ are $e^{\pm 2i\theta}$). Expressing the constraint
\eqref{C} in terms of 
\begin{equation}
  \begin{split}
    D &= \W\Wd = X^2+Y^2-i[X,Y]\\
    \Dt &= \Wd\W = X^2+Y^2+i[X,Y],
  \end{split}
\end{equation}
giving
\begin{align}
  \paraa{D+\Dt-2\mu}^2+\frac{\paraa{D-\Dt}^2}{\tan^2\theta}=4\cdot\mid,\label{CDDt}
\end{align}
one sees that the transformation \eqref{Matrixiteration} must leave
invariant the ellipse given via \eqref{CDDt}. Although this becomes
obvious in the ``circle-coordinates''
\begin{equation}
  \begin{split}
    Z &= \frac{D-\Dt}{2\h}=-\frac{i}{\h}[X,Y]\\
    \varphi &= \frac{1}{2}\paraa{D+\Dt}-\mu = X^2+Y^2-\mu,
  \end{split}
\end{equation}
in which the transformation \eqref{xvn} between eigenvalue-pairs is simply a
rotation by $2\theta$,
\begin{align}
  \begin{pmatrix}
    z_{n+1}\\\varphi_{n+1}
  \end{pmatrix}
  =
  \begin{pmatrix}
    \cos(2n\theta) & -\sin(2n\theta)\\
    \sin(2n\theta) & \cos(2n\theta)
  \end{pmatrix}
  \begin{pmatrix}
    z_1 \\ \varphi_1
  \end{pmatrix},
\end{align}
the picture of a by $45^\circ$ rotated ellipse (with halfaxes 1 and
$\h=\tan\theta$) lying in the $(d,\dt)$-plane is extremely useful,
 in particular when discussing the $(\mu,\theta)$-dependence of $N$-dimensional representations of
eqs  \eqref{XY}--\eqref{C} (s.b.).

Remembering, e.g. that $D=\W\Wd$ we get
\begin{align*}
  |W_k|^2 = d_k,
\end{align*}
which is only solvable if the ellipse defined by $\xv_1=(d,\dt)$
entirely lies in the first quadrant. This observation leads to the
following: Assume that $\theta=\pi/N$ for some $N>0$ and let
$\xv=(d,\dt)$ lie on the ellipse $(d+\dt-2\mu)^2+(d-\dt)^2/\h^2=4$. If
$\mu\leq1$ then there exists no $\xv=(d,\dt)$ such that $d_n>0$ and
$\dt_n>0$ for $n=1,2,\ldots,N$. If $\mu>1$ and $\cos\theta>1/\mu$ then,
for every choice of $\xv=(d,\dt)$ on the ellipse, $d_n>0$ and $\dt_n>0$
for all $n\geq 1$.

\begin{figure}[h]
\begin{center}
  \includegraphics*[height=4cm,width=4cm]{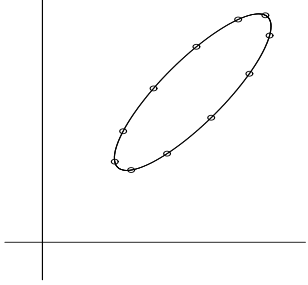}
  \caption{$\mu=2$, $N=11$.}\label{fig:mu2N11}
\end{center}
\end{figure}

\noindent These solutions take the form
\begin{align}
\begin{split}
  X &= \frac{1}{2} 
  \begin{pmatrix}
    0      & x_1    & 0      & \cdots & 0      & x_N \\
    x_1    & 0      & x_2    & \cdots & 0      & 0   \\
    0      & x_2    & 0      & \ddots & 0      & 0   \\
    \vdots & \ddots & \ddots & \ddots & \ddots & \vdots\\
    0      & 0      & \cdots & x_{N-2}& 0      & x_{N-1}\\
    x_N    & 0      & \cdots & 0      & x_{N-1}& 0
  \end{pmatrix}\\
  Y &=-\frac{i}{2}
  \begin{pmatrix}
    0      & y_1    & 0      & \cdots & 0      & -y_N \\
    -y_1    & 0      & y_2    & \cdots & 0      & 0   \\
    0      & -y_2    & 0      & \ddots & 0      & 0   \\
    \vdots & \ddots & \ddots & \ddots & \ddots & \vdots\\
    0      & 0      & \cdots & -y_{N-2}& 0      & y_{N-1}\\
    y_N    & 0      & \cdots & 0      & -y_{N-1}& 0
  \end{pmatrix}\\
  Z &= \diag\paraa{z_1,z_2,\ldots,z_N}\\
  x_l &= y_l = 
  \sqrt{\mu+\frac{\cos\paraa{\frac{2\pi
  l}{N}+\beta}}{\cos\paraa{\pi/N}}}=w_l\\
  z_l &= -\sin\parac{\frac{2\pi l}{N}-\frac{\pi}{N}+\beta}
\end{split}
\end{align}
and the ellipse, on which $(d_i,\dt_i)$ lie, will typically look like
in Figure \ref{fig:mu2N11}.

\begin{figure}[h]
\begin{center}
  \includegraphics*[height=4cm,width=4cm]{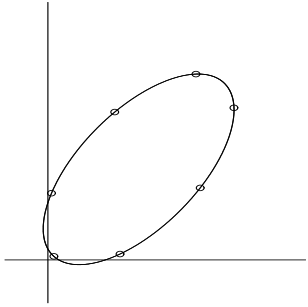}
  \caption{$\mu\approx 1.055$, $N=7$.\label{fig:2}}
\end{center}
\end{figure}

\noindent In the region $1<\mu<1/\cos\theta$ the set of $(d,\dt)$, for which
$\dn,\dtn>0$ for all $n$, is a union of disjoint intervals whose
lengths decreases and eventually become a set of $N$
distinct points as $\mu\to 1$ (however, making $d^{(i)}=0$ for some
$i$, giving not a ``loop'', but a ``string'' solution. This transition
will be discussed in Section 6). For a loop solution in this region,
the points $(d_i,\dt_i)$ will precisely ``miss'' the negative
region, like in Figure \ref{fig:2}.
\noindent For $\mu<1$, by making a ``string'' Ansatz for $\W$
\begin{align}
  W = 
  \begin{pmatrix}
    0 & w_1 & 0   & \cdots & 0 \\
    0 & 0   & w_2 & \cdots & 0 \\
    \vdots & \vdots & \ddots & \ddots & \vdots \\
    0 & 0 & \cdots & 0 & w_{N-1} \\
    0 & 0 & \cdots & 0 & 0
  \end{pmatrix},\label{Wstring}
\end{align}
one derives, from $D=\W\Wd$ and $\Dt=\Wd\W$, the condition
$\dt_1=0=d_N$, which makes the choice
$d_1=2\sin\theta\paraa{\mu\sin\theta+\sqrt{1-\mu^2\cos^2\theta}}$
necessary.  Can we now, for given $N$, find $\theta$ such that
$d_N=0$? Assume that $-1<\mu<1/\cos\theta$ and let
$\xv_1=\paraa{2\sin\theta\paraa{\mu\sin\theta+\sqrt{1-\mu^2\cos^2\theta}},0}$.
If $\theta$ is a solution of
\begin{align}
  \frac{\cos(N\theta)}{\cos\theta}=-\mu,
\end{align}
then $d_n=0$ and $d_n,\dt_n>0$ for $n=2,3,\ldots,N-1$.

There are three values of $\mu$ for which it is particularly easy to calculate
these solutions explicitly: $\mu=1,\mu=\h$ and $\mu=0$.

\begin{figure}[h]
  \begin{center}
    \includegraphics*[height=4cm,width=4cm]{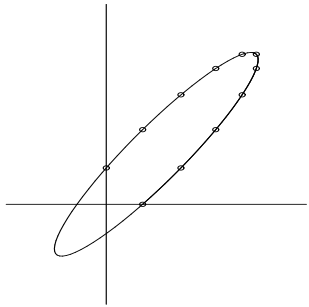}
    \caption{$\mu=1/2$, $N=11$ and $\theta\approx 0.189$.}\label{fig:muhalf}
  \end{center}
\end{figure}

\noindent\underline{$\mu=1$:}
\begin{align}
\begin{split}
  X &=\frac{1}{2}
  \begin{pmatrix}
    0      & x_1    & 0   & \cdots & 0\\
    x_1    & 0      & x_2 & \cdots & 0\\
    \vdots & \ddots & \ddots & \ddots & \vdots\\
    0      & 0      & x_{N-2}      & 0      & x_{N-1}\\
    0      & 0      & 0      & x_{N-1}         & 0
  \end{pmatrix},\quad
  Y = -\frac{i}{2}
  \begin{pmatrix}
    0      & y_1    & 0   & \cdots & 0\\
    -y_1    & 0      & y_2 & \cdots & 0\\
    \vdots & \ddots & \ddots & \ddots & \vdots\\
    0      & 0      & -y_{N-2}      & 0      & y_{N-1}\\
    0      & 0      & 0      & -y_{N-1}         & 0
  \end{pmatrix}\\
  Z &= \diag\paraa{z_1,z_2,\ldots,z_N},\quad  z_l =\sin\parac{\frac{2\pi l}{N+1}}\\
  x_l&=y_l=
  \sqrt{1-\frac{\cos\parab{\frac{(2l+1)\pi}{N+1}}}
    {\cos\frac{\pi}{N+1}}}
\end{split}
\end{align}
\noindent\underline{$\mu=\h=\tan\frac{\pi}{2(N-1)}$:}
\begin{align}
\begin{split}
  X &= \frac{1}{2}
  \begin{pmatrix}
    0      & x_1    & 0   & \cdots & 0\\
    x_1    & 0      & x_2 & \cdots & 0\\
    \vdots & \ddots & \ddots & \ddots & \vdots\\
    0      & 0      & x_{N-2}      & 0      & x_{N-1}\\
    0      & 0      & 0      & x_{N-1}         & 0
  \end{pmatrix},\quad
  Y = -\frac{i}{2}
  \begin{pmatrix}
    0      & y_1    & 0   & \cdots & 0\\
    -y_1    & 0      & y_2 & \cdots & 0\\
    \vdots & \ddots & \ddots & \ddots & \vdots\\
    0      & 0      & -y_{N-2}      & 0      & y_{N-1}\\
    0      & 0      & 0      & -y_{N-1}         & 0
  \end{pmatrix}\\
  Z &= \diag\paraa{z_1,z_2,\ldots,z_N},\quad  z_l = \cos\parac{\frac{(l-1)\pi}{N-1}}\\
  x_l&=y_l=
  \sqrt{\frac{2}{\cos\frac{\pi}{2(N-1)}}}
  \sqrt{\sin\parac{\frac{\pi l}{2(N-1)}}
    \cos\parac{\frac{\pi(l-1)}{2(N-1)}}}
\end{split}
\end{align}
\noindent\underline{$\mu=0$:}
\begin{align}
  \begin{split}
  X &= \frac{1}{2}
  \begin{pmatrix}
    0      & x_1    & 0   & \cdots & 0\\
    x_1    & 0      & x_2 & \cdots & 0\\
    \vdots & \ddots & \ddots & \ddots & \vdots\\
    0      & 0      & x_{N-2}      & 0      & x_{N-1}\\
    0      & 0      & 0      & x_{N-1}         & 0
  \end{pmatrix},\quad
  Y = -\frac{i}{2}
  \begin{pmatrix}
    0      & y_1    & 0   & \cdots & 0\\
    -y_1    & 0      & y_2 & \cdots & 0\\
    \vdots & \ddots & \ddots & \ddots & \vdots\\
    0      & 0      & -y_{N-2}      & 0      & y_{N-1}\\
    0      & 0      & 0      & -y_{N-1}         & 0
  \end{pmatrix}\\
  Z &= \diag\paraa{z_1,z_2,\ldots,z_N},\quad z_l = \cos\parac{\frac{l\pi}{N}-\frac{\pi}{2N}}\\
  x_l&=y_l=
  \sqrt{
    \frac{1}{\cos\frac{\pi}{2N}}\sin\parab{\frac{l\pi}{N}}
  }
  \end{split}
\end{align}
In the region $-1<\mu\leq 1$ the corresponding ellipse for a string
solution will typically look like Figure \ref{fig:muhalf}.

\noindent Let us now derive the subcritical condition for existence of
an $N$-dimensional string representation,
\begin{align}
  \cos(N\theta)+\mu\cos\theta=0,
\end{align}
valid for all $\mu\in(-1,1/\cos\theta]$; with no solution for $\mu=-1$ as
  $(X^2+Y^2+1)^2+Z^2=0$ cannot have any nontrivial solutions.  It is
  useful to remember that
\begin{align}
  \begin{split}
    \mu &= 1\quad\Rightarrow\quad \theta=\frac{\pi}{N+1}\\
    \mu &= 0\quad\Rightarrow\quad \theta=\frac{\pi}{2N}\\
    \mu &= \h\quad\Rightarrow\quad \theta=\frac{\pi}{2(N-1)}.
  \end{split}
\end{align}
For $\mu\in[\tan\theta,1]$ (other values of $\mu$ can be
treated analogously):
\begin{equation}
  \begin{split}
    d_1&=2\mu\sin\theta\parac{\sin\theta+\sqrt{\frac{1}{\mu^2}-\cos^2\theta}}
    \quad,\quad \dt_1=0\\
    d_N&=0\quad,\quad \dt_N=d_1
  \end{split}
\end{equation}
gives
\begin{equation}
  \begin{split}
    z_1&=\frac{\cot\theta}{2}d_1=\mu\cos\theta\parac{\sin\theta+\sqrt{\frac{1}{\mu^2}-\cos^2\theta}}=-z_N\\
    \varphi_1&=\frac{d_1}{2}-\mu=\mu\parac{\sin\theta\sqrt{\frac{1}{\mu^2}-\cos^2\theta}-\cos^2\theta}=\varphi_N<0.
  \end{split}
\end{equation}
Let $\chi$ be the angle from $(z_1,\varphi_1)$ to
$(z_N,\varphi_N)$: 
\begin{figure}[h]
\begin{center}
\setlength{\unitlength}{0.00087489in}
\begingroup\makeatletter\ifx\SetFigFont\undefined%
\gdef\SetFigFont#1#2#3#4#5{%
  \reset@font\fontsize{#1}{#2pt}%
  \fontfamily{#3}\fontseries{#4}\fontshape{#5}%
  \selectfont}%
\fi\endgroup%
{\newcommand{\dashlinestretch}{30}
\begin{picture}(3032,2759)(0,-10)
\put(1286.948,1350.021){\arc{534.795}{2.3877}{7.0897}}
\path(986.067,1204.874)(1092.000,1167.000)(1098.697,1279.300)
\thicklines
\put(1282,1327){\ellipse{1548}{1548}}
\put(692,822){\blacken\ellipse{80}{80}}
\put(692,822){\ellipse{80}{80}}
\put(1867,827){\blacken\ellipse{80}{80}}
\put(1867,827){\ellipse{80}{80}}
\path(22,1327)(2947,1327)
\blacken\thinlines
\path(2707.000,1282.000)(2947.000,1327.000)(2707.000,1372.000)(2707.000,1282.000)
\thicklines
\path(1282,1327)(697,832)
\path(1282,1327)(697,832)
\blacken\thinlines
\path(1327.000,2482.000)(1282.000,2722.000)(1237.000,2482.000)(1327.000,2482.000)
\path(1282,2722)(1282,22)
\thicklines
\path(1287,1322)(1872,827)
\path(1287,1322)(1872,827)
\put(1462,2562){\makebox(0,0)[lb]{\smash{{{\SetFigFont{12}{14.4}{\rmdefault}{\mddefault}{\updefault}$\varphi$}}}}}
\put(3032,1292){\makebox(0,0)[lb]{\smash{{{\SetFigFont{12}{14.4}{\rmdefault}{\mddefault}{\updefault}$z$}}}}}
\put(1387,1667){\makebox(0,0)[lb]{\smash{{{\SetFigFont{12}{14.4}{\rmdefault}{\mddefault}{\updefault}$\chi$}}}}}
\put(92,522){\makebox(0,0)[lb]{\smash{{{\SetFigFont{12}{14.4}{\rmdefault}{\mddefault}{\updefault}$(z_N,\varphi_N)$}}}}}
\put(1927,527){\makebox(0,0)[lb]{\smash{{{\SetFigFont{12}{14.4}{\rmdefault}{\mddefault}{\updefault}$(z_1,\varphi_1)$}}}}}
\end{picture}
}
\caption{\label{fig:3}}
\end{center}
\end{figure}
If we, by doing $(N-1)$ rotations by $2\theta$,
want to go from $(z_1,\varphi_1)$ to $(z_N,\varphi_N)$, we get the
condition
\begin{align}
  2\theta(N-1)=\pi+2\arccos z_1\label{NTone}
\end{align}
since $\mu\in[\tan\theta,1]$. Rearranging \eqref{NTone}, and inserting
the expression for $z_1$ one obtains, by taking
$\cos$ of both sides
\begin{align}
  \sin(N-1)\theta-\mu\cos\theta\sin\theta = \cos\theta\sqrt{1-\mu^2\cos^2\theta}.
\end{align}
Squaring the expression and solving for $\mu$ yields
\begin{align}
  \begin{split}
    \mu &= \tan\theta\sin(N-1)\theta\pm\sqrt{1-\sin^2(N-1)\theta}\\
    &= \mp\frac{\cos\paraa{(N-1\pm 1)\theta}}{\cos\theta}.
  \end{split}
\end{align}
By knowing the sign of $\cos(N\theta)$ and $\cos\theta$, we see that one of the roots
is a false root, leaving
\begin{align}
  \mu = -\frac{\cos(N\theta)}{\cos\theta}.
\end{align}
For given $\mu$, out of the solutions for $\theta$,
it is only the smallest that gives
the string solution; the larger $\theta$'s correspond to a total
rotation of more than $2\pi$ (giving negative $d_i$'s).

\section{The singularity}\label{sec:sing}

\begin{figure}[h]
\begin{center}
\includegraphics[height=7cm]{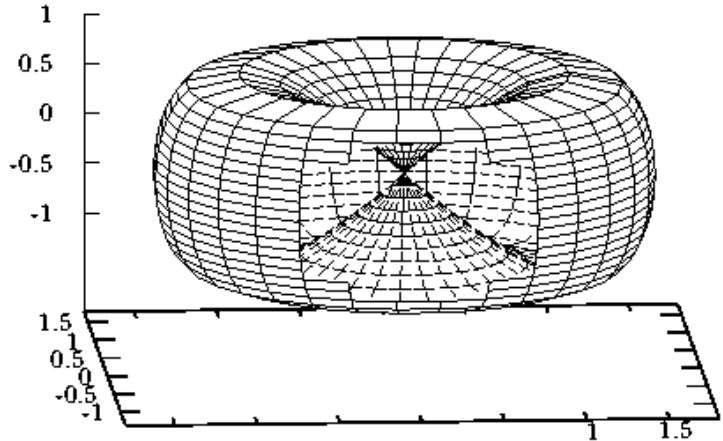}
\caption{\label{fig:1}$(x^2+y^2-1)^2+z^2=1$}
\end{center}
\end{figure}

\noindent The equation
\begin{equation}\label{eq:sing_1}
(x^2+y^2-\mu)^2+z^2=1
\end{equation}
defines a regular surface, except for $\mu=1$ (the singularity being
at the origin, $x=y=z=0$, see figure \ref{fig:1}). How is this
singularity reflected in the ``fuzzy world'', \ie when looking at
finite dimensional representation of (\ref{XY}-\ref{C})? Interestingly,
$\Sigma_{\hbar=\tan\theta}(P=x^2-1)$ ``does exist'', for $\theta$ an
integer fraction of $\pi$. Equations (\ref{XY}-\ref{C}) do have, for
$\theta=\pi/N$, (up to conjugation) a unique (``irreducible'') $N-1$
dimensional solution:
\begin{eqnarray}
X=\frac{1}{2}\left(\begin{matrix} 0 & x_2 & & \\ x_2 & 0 & \ddots & \\
& \ddots & \ddots & x_{N-1} \\ & & x_{N-1} & 0
\end{matrix}\right),\qquad Y=-\frac{i}{2}\left(\begin{matrix}
0 & y_2 &  & \\ 
-y_2 & 0 & \ddots & \\ 
 & \ddots & \ddots & y_{N-1} \\ 
 & & -y_{N-1} & 0
\end{matrix}\right),\nonumber\\
Z=\left(\begin{matrix}
z_2 & & & 0\\ 
 & z_3 & 0& \\ 
 & 0& \ddots & \\ 
0 & & & z_{N-1}
\end{matrix}\right)\label{eq:sing_2}\\
x_l=y_l=\sqrt{1-\frac{\cos\left(\frac{2\pi
l}{N}-\frac{\pi}{N}\right)}{\cos\left(\frac{\pi}{N}\right)}},\quad
z_l=\sin\left(\frac{2\pi(l-1)}{N}\right),\quad\text{for}\
l=2,3,\ldots,N.\nonumber
\end{eqnarray}
What happens, is the following: while for large enough $\mu$
($\mu>1/\cos\theta>1$), the ellipse lies entirely in the first
quadrant of the $(d,\tilde{d})$ plane ($d,\tilde{d}>0$), leading to an
$N$-dimensional representation (if $N\theta=2\pi$) for arbitrary
initial conditions (\ie an arbitrary initial point on the ellipse),
this is no longer the case when $\mu$ approaches 1; when $\mu$ becomes
smaller than $1/\cos\theta$, and approaches $1$ from above, the
continuous range of inital conditions gradually shrinks, leaving at
$\mu=1$ precisely \emph{one} ``$N$-dimensional'' representation, which
however has the additional feature that in the limit ($\mu=1$) the
first row and column of $X$, $Y$ and $Z$ becomes identically
zero. This drop of the dimensionality (by 1) for given $\theta=\pi/N$
could be viewed as reflecting the singularity, while on the other
hands it cleverly (``smoothly'') leads over to the subcritical
($\mu<1$) $N$-dependence of $\theta$.\\ The only representation that
survives as $\mu\rightarrow 1$ is the one given by
\begin{align}
\begin{split}
X&=\frac{1}{2}\left(\begin{matrix}
0 & x_1 & 0 & x_N\\ 
x_1 & 0 & \ddots & 0\\ 
 0 & \ddots & \ddots & x_{N-1} \\ 
 x_N & 0 & x_{N-1} & 0
\end{matrix}\right),\qquad Y=-\frac{i}{2}\left(\begin{matrix}
0 & y_1 & 0  & -y_N\\ 
-y_1 & 0 & \ddots & 0\\ 
 0 & \ddots & \ddots & y_{N-1} \\ 
 y_N & 0 & -y_{N-1} & 0
\end{matrix}\right),\\ 
Z&=\diag(z_1,z_2,\ldots,z_N)\label{eq:sing_3}\\
x_l&=y_l=\sqrt{\mu-\frac{\cos\left(\frac{2\pi
l}{N}-\frac{\pi}{N}\right)}{\cos\left(\frac{\pi}{N}\right)}},\quad
z_l=\sin\left(\frac{2\pi(l-1)}{N}\right),\quad\text{for}\
l=1,2,\ldots,N.
\end{split}
\end{align}
\ie $d_1=\tilde{d_1}=\mu-1$ (the lower tip of the ellipse) as the
starting point (due to $\tilde d_2=d_1=\mu-1>0$ it is qualitatively
clear, that, with that initial condition one ``jumps'' over the small
region of negative $\tilde d$, as well as, ``at the end'' the one of
negative $d$).\\ The drop in dimensionality (for $\mu=1$) then is
simply the vanishing of $x_1=y_1$ and $x_N=y_N$. One could of course
relabel the points (always start with the second point, instead of the
lower tip) such that $\mu=1$ the upper $(N-1)\times (N-1)$ block has a
smooth limit (and the $N^{\operatorname{th}}$ row/column
``disappears'' as $\mu\rightarrow 1$). As noted above, the subcritical
behaviour of $\theta$ as a function of $\mu$ (to have a
$N$-dimensional representation) is more involved. As the allowed
part of the ellipse shrinks to zero as $\mu$ goes from $1$ to $-1$),
$\theta$ has to accordingly decrease (for fixed dimension $N$ of the
representation); for $\mu=0$, it is equal to $\pi/2 N$.\\ As a
reflection of the classical singularity at $\mu=1$, the quantum
(fuzzy) analogue manifests itself not only (by the sudden drop in
dimension) at $\mu=1$, but also in the neighbouring region,
$1<\mu<1/\cos\theta$, which we shall now discuss in detail: in this
range,
\begin{equation}
d_{\pm} = 2\sin\theta\left(\mu\sin\theta\pm\sqrt{1-\mu^2\cos^2\theta}\right)\,;\label{eq:sing_4}
\end{equation}
the $d$-values at which the ellipse crosses the $d$-axis, are both
(real and) positive. The corresponding points on the $z$--$\varphi$
circle have coordinates
\begin{eqnarray}
z_\pm = \cos\theta\left(\mu\sin\theta\pm\sqrt{1-\mu^2\cos^2\theta}\right)>0\nonumber\\
\varphi_\pm = \pm\sin\theta\sqrt{1-\mu^2\cos^2\theta}-\mu\cos^2\theta <0.\label{eq:sing_5}
\end{eqnarray}
Let us denote the angles between the negative $\varphi$ axis and the 2
points (given by (\ref{eq:sing_5})) by $\psi_\pm$ (cp. Figure \ref{fig:2}).\\ At
$\mu=1/\cos\theta$: $\psi_+=\psi_-=\theta$ and at $\mu=1$:
$\psi_+^{(1)}=2\theta$, $\psi_-^{(1)}=0$. To find a ``closed string''
solution, initial conditions with angles
$\psi\in\left(\psi_-,\psi_+\right)$, are forbidden (as well as those
regions obtained by rotating the interval $\left(\psi_-,\psi_+\right)$
by $2k\theta$ ($k=1,\ldots,N-1$). To the ``black'' regions one has to
add rotation images of the corresponding part of the ellipse that
extended into the negative $d$-region
$\psi\in\left(\tilde\psi_+,\tilde\psi_-\right)=\left(-\psi_+,-\psi_-\right)$.\\
Hence one obtains
\begin{equation}
B:=\bigcup_{k=0}^{N-1}\left(-\psi_++\frac{2\pi}{N}+\frac{2\pi}{N}k,\psi_++\frac{2\pi}{N}k\right)\label{eq:sing_6}
\end{equation}
as the forbidden (``black'') region. $B$ continuously grows from empty (at $\mu=1/\cos\theta$) to
\begin{equation}
B_1 = [0,2\pi)\setminus\left\{0,\frac{2\pi}{N},\frac{4\pi}{N},\ldots,2\pi\frac{N-1}{N}\right\}\label{eq:sing_7}
\end{equation}
(at $\mu=1$, where $\psi_+=2\theta=2\pi/N$; due to $d(\psi=0)=0=\tilde
d(\psi=0)$ at $\mu=1$ the ``closed string'' solution disappears, the
corresponding dimensionality drops by 1, and the ``truely
$N$-dimensional'' open-string representation then corrponds to
$\theta=\pi/(N-1)$).\\ Note that while in the ``critical region''
($1\leq\mu\leq 1/\cos\theta$) both --closed and open-- string
$N$-dimensional solutions exist, they never coexist for the same value
of $\theta$ (\resp $\hbar$). $N$-dimensional closed-string-solutions
naturally require $\theta=2\pi/N$, while $N$-dimension
open-string-solutions are subject to the quantisation condition
\begin{equation}
\cos(N\theta)+\mu\cdot\cos\theta=0\label{eq:sing_8}
\end{equation}
(the derivation is identical to the one for the subcritical region(s),
$\mu<1$), which gives $\theta=\pi/N$ for $\mu=1/\cos\theta$, and
$\theta=\pi/(N-1)$ for $\mu=1$ (and no integer lying between $N$ and
$N-1$).

\section*{Acknowledgement}

We would like to thank the Royal Institute of Technology, the Albert
Einstein Institute, the Japan Society for the Promotion of Science,
the Sonderforschungsbereich "Raum-Zeit-Materie", and the Marie Curie
Research Training Network ENIGMA for financial support and
hospitality.


\end{document}